# Quantum Annealing for Combinatorial Optimization: A Benchmarking Study

**Authors**: Seongmin Kim[1,4], Sang-Woo Ahn[2], In-Saeng Suh[4], Alexander W. Dowling[3,*], Eungkyu Lee[2,*], and Tengfei Luo[1,*]

[1]Department of Aerospace and Mechanical Engineering, University of Notre Dame; Notre Dame, Indiana 46556, United States.
[2]Department of Electronic Engineering, Kyung Hee University; Yongin-Si, Gyeonggi-do 17104, Republic of Korea.
[3]Department of Chemical and Biomolecular Engineering, University of Notre Dame; Notre Dame, Indiana 46556, United States.
[4]National Center for Computational Sciences, Oak Ridge National Laboratory, Oak Ridge, Tennessee 37830, United States.
*Corresponding author. Email: adowling@nd.edu, eleest@khu.ac.kr, and tluo@nd.edu

Quantum annealing (QA) has the potential to significantly improve solution quality and reduce time complexity in solving combinatorial optimization problems compared to classical optimization methods. However, due to the limited number of qubits and their connectivity, the QA hardware did not show such an advantage over classical methods in past benchmarking studies. Recent advancements in QA with more than 5,000 qubits, enhanced qubit connectivity, and the hybrid architecture promise to realize the quantum advantage. Here, we use a quantum annealer with state-of-the-art techniques and benchmark its performance against classical solvers. To compare their performance, we solve over 50 optimization problem instances represented by large and dense Hamiltonian matrices using quantum and classical solvers. The results demonstrate that a state-of-the-art quantum solver has higher accuracy (~0.013%) and a significantly faster problem-solving time (~6,561×) than the best classical solver. Our results highlight the advantages of leveraging QA over classical counterparts, particularly in hybrid configurations, for achieving high accuracy and substantially reduced problem solving time in large-scale real-world optimization problems.

## Introduction

Quantum computers mark a paradigm shift to tackle challenging tasks that classical computers cannot solve in a practical timescale[1,2]. The quantum annealer is a special quantum computer designed to solve combinatorial optimization problems with problem size-independent time complexity[3-5]. This unique quantum annealing (QA) capability is based on the so-called adiabatic process[6,7]. During this process, entangled qubits naturally evolve into the ground state of a given Hamiltonian to find the optimal vector of binary decisions for the corresponding quadratic unconstrained binary optimization (QUBO) problem[8-10]. The adiabatic theorem of quantum mechanics ensures that QA identifies the optimal solution regardless of the size and landscape of the combinatorial parametric space, highlighting QA as a powerful and practical solver[11-14]. The ability to efficiently explore high-dimensional combinational spaces makes QA capable of handling a wide range of optimization tasks[4,5,10,15,16].

The potential merit of QA motivates the systematic comparison with classical counterparts (e.g., simulated annealing, integer programming, steepest descent method, tabu search, and parallel tempering with isoenergetic cluster moves), focusing on the solution quality and the time complexity. While previous benchmarking studies showed some advantages of QA, most used low-dimensional or the sparse configuration of QUBO matrices due to the lack of available qubits in the QA hardware and poor topology to connect qubits[17-19]. For example, O'Malley et al. [17] compared the performance of QA with classical methods (mathematical programming), but they limited the number of binary variables to 35 due to the QA hardware limitation. Similarly, Tasseff et al. [18] highlighted the potential advantages of QA compared to classical methods (such as simulated annealing, integer programming, and Markov chain Monte Carlo) for sparse optimization problems containing up to 5,000 decision variables and 40,000 quadratic terms. Haba et al. [19] demonstrated that a classical solver (integer programming) could be faster than QA for small problems, e.g., ~100 decision variables. Consequently, these benchmarking studies show that QA methods and their classical counterparts can exhibit similar solution quality and time complexity. However, such low-dimensional or sparse QUBOs considered in the previous benchmarking studies are challenging to map to a wide range of practical problems, which usually require high-dimensional and dense configuration of QUBO matrices[4,5,10,20]. For example, in our previous QA optimization of one-dimensional and two-dimensional optical metamaterials, the QUBO matrices exhibit these properties (Fig. S1) [4,5,16,20].

The state-of-the-art QA hardware (D-Wave Advantage System) features more than 5,000 qubits, advanced topology to connect qubits, and efficient hybrid algorithms (e.g., Leap Hybrid sampler). For example, the recent development (e.g., Pegasus topology) has increased qubit connectivity from 6 to 15[21-23]. Improved qubit connectivity reduces the need for complex embedding processes, which map problem variables to physical qubits on the hardware. With better connectivity, such as in D-Wave's Pegasus topology, the embedding process becomes more efficient and can better preserve the structure of dense optimization problems. This enhancement allows the quantum annealer to increase the potential for finding high-quality solutions[24,25]. In addition, a QUBO decomposition algorithm (i.e., QBSolv) splits a large QUBO matrix into small pieces of sub-QUBO matrices, allowing us to handle a QUBO matrix with dimensions higher than the maximum number of qubits in the QA hardware[26,27]. Given these advancements, it is imperative to study the performance of the state-of-the-art QA system for high-dimensional and dense configuration of QUBO matrices, and systemically compare solution quality and the time complexity with the classical counterparts.

In this work, we benchmark the performance of quantum solvers against classical algorithms in solving QUBO problems with large and dense configurations to represent real-world optimization problems. We analyze the solution quality and the required time to solve these benchmark problems using several quantum and classical solvers. This benchmarking study provides important insights into employing QA in practical problem-solving scenarios.

## Results

We present a benchmarking study on combinatorial optimization problems representing real-world scenarios, e.g., materials design, characterized by dense and large QUBO matrices (Fig. S1). These problems are non-convex and exhibit a highly complex energy landscape, making it challenging and time-consuming to identify accurate solutions. Classical solvers, such as integer programming (IP), simulated annealing (SA), steepest descent (SD), tabu search (TS), parallel tempering with isoenergetic cluster moves (PT-ICM), perform well for small-scale problems. However, they are often relatively inaccurate for larger problems (problem size ≥ 1,000; Fig. 1a). In particular, SD and TS show low relative accuracy compared to other solvers. The combination of PT and ICM leverages the strengths of both techniques: PT facilitates crossing energy barriers, while ICM ensures exploration of the solution space, effectively covering broad and diverse regions. This makes PT-ICM particularly effective for exploring complex optimization spaces and enhancing convergence toward the global optimum[46,47]. However, the performance of PT-ICM can be problem-dependent[48]. While it can work well for sparse problems, its effectiveness decreases for denser problems[46]. Consequently, although SA, and PT-ICM perform better than SD and TS, they also fail to find high-quality solutions for large-scale problems.

To address these limitations, QUBO decomposition strategies can be employed to improve the relative accuracy. For example, integrating QUBO decomposition with classical solvers (e.g., SA–QBSolv and PT-ICM–QBSolv) improves their performance. Nonetheless, these approaches often remain insufficient for handling massive problems effectively, particularly considering problem-solving time (Fig. 1b), which will be further discussed in the following. On the other hand, quantum solvers provide excellent performance for solving these dense and large-scale problems representing real-world optimization scenarios. Although QA can perform excellently for small problems, it has difficulty solving large and dense QUBOs due to the limited number of qubits (5,000+) and connectivity (15). Several prior studies reported that QA may not be efficient since it cannot effectively handle dense and large QUBOs due to hardware limitations[23,53,54]. However, when it runs with the QUBO decomposition strategy (i.e., QA–QBSolv), large-scale problems ($n \geq 100$) can be effectively handled. Furthermore, hybrid QA (HQA), which integrates quantum and classical approaches, also can solve large-scale problems efficiently. As a result, the quantum solvers consistently identify high-quality solutions across all problem sizes (Fig. 1a).

Computational time is also a critical metric for evaluating solver performance. Classical solvers exhibit rapidly increasing solving times as problem sizes grow, making them impractical for large-scale combinatorial optimization problems (Fig. 1b). While SD and TS are faster than other classical solvers, their relative accuracies are low, as can be seen in Fig. 1a. It is worth noting that the SA, and PT-ICM solvers struggle to handle problems with more than 3,000 variables due to excessively long solving time or computational constraints (e.g., memory limits). Although the IP solver is faster than SA and PT-ICM, its solving time increases greatly with problem size. The QUBO decomposition strategy significantly reduces computational time, yet quantum solvers remain faster than their classical counterparts across all problem sizes. For instance, for a problem size of 5,000, the solving time for HQA is 0.0854 s and for QA–QBSolv is 74.59 s, compared to 167.4 s and 195.1 s for SA–QBSolv and PT-ICM–QBSolv, respectively, highlighting superior efficiency of the quantum solvers.

To further evaluate scalability, we conduct a systematic benchmarking study on QUBO problems (size: up to 10,000 variables), designed to mimic real-world scenarios through randomly generated elements. PT-ICM is excluded from this analysis due to excessive solving times compared to other solvers (Fig. 1b). As shown in Fig. 2, classical solvers (IP, SA, SD, and TS) are accurate for smaller problems but become inaccurate as the problem size increases. Consistent with the results in Fig. 1, the SD and TS solvers exhibit low relative accuracy even for a relatively small problem (e.g., 2,000). IP and SA are more accurate than SD and TS but fail to identify the optimal state for large problems. It is known that IP can provide global optimality guarantees[40], but our study highlights that proving a solution is globally optimal is challenging for large and dense problems. For example, in one case ($n = 7,000$), the optimality gap remains as large as ~17.73%, where the best bound is -19,660 while the solution obtained from the IP solver is -16,700, with the optimality gap not narrowing even after 2 hours of runtime. The relative accuracy can be improved by employing the QUBO decomposition strategy (e.g., SA–QBSolv), yet it still fails to identify high-quality solutions for problem sizes exceeding 4,000. In contrast, quantum solvers demonstrate superior accuracy for large-scale problems. Notably, the HQA solver consistently outperforms all other methods, reliably identifying the best solution regardless of problem size (Fig. 2).

Fig. 3a shows that the solving time rapidly increases as the problem size increases for the classical solvers, indicating that solving combinatorial optimization problems with classical solvers can become intractable for large-size problems (Fig. 3b). The solving time trends with increasing problem size agree well with the theoretical time complexities of the classical solvers (Fig. 3b and Fig. S3, see 2-4-2. Computational Time section). While the IP solver can be faster than other classical solvers, it also requires significant time for large problems (e.g., $n > 5,000$). The use of the QUBO decomposition strategy dramatically reduces the solving time, but the quantum solvers consistently outpace classical counterparts (Fig. 3a). For example, the solving time ($n = 10,000$) is 0.0855 s for HQA, 101 s for QA–QBSolv, and 561 s for SA–QBSolv.

Decomposing a large QUBO into smaller pieces leads to a higher relative accuracy, as a solver can find better solutions for each decomposed QUBOs, mitigating the current hardware limitations. Note that the accuracy of QA for QUBOs with problem sizes of 30 and 100 is, respectively, 1.0 and 0.9956 (without leveraging the QUBO decomposition method). Hence, the accuracy of QA–QBSolv with a sub-QUBO size of 30 is higher than that with a sub-QUBO size of 100, as decomposed QUBOs with a smaller size fit the QA hardware better (Fig. 4a). However, a smaller sub-QUBO size results in a greater number of sub-QUBOs after decomposition, leading to increased time required to solve all decomposed problems (Fig. 4b). It is noted that the QA–QBSolv solver does not guarantee finding the best solution for large problems (size > 4,000), resulting in lower accuracies regardless of sub-QUBO sizes, as can be seen in Fig. 2 and Fig. 4a.

Our results show that HQA, which incorporates QA with classical algorithms to overcome the current quantum hardware limitations, is currently the most efficient solver for complex real-world problems that require the formulation of dense and large QUBOs. In this context, we define "Quantum Advantage" as the ability of a quantum-enhanced solver to achieve high accuracy and significantly faster problem-solving time compared to the classical solvers for large-scale optimization problems. Our findings suggest that leveraging quantum resources, particularly in hybrid configurations, can provide a computational advantage over classical approaches. Besides, as the current state of HQA demonstrates, we expect QA will have much higher accuracy and require much shorter time to solve QUBO problems with the development of the quantum hardware with more qubits and better qubit connectivity.

## Discussion

This work comprehensively compares state-of-the-art QA hardware and software against several classical optimization solvers for large and dense QUBO problems (up to 10,000 variables, fully connected interactions). The classical solvers struggled to solve large-scale problems, but their performance can be improved when combined with the QUBO decomposition method (i.e., QBSolv). Nevertheless, they become inaccurate and inefficient with increasing problem size, indicating that classical methods can face challenges for complex real-world problems represented by large and dense QUBO matrices. On the contrary, HQA performs significantly better than its classical counterparts, exhibiting the highest accuracy (~0.013% improvement) and shortest time to obtain solutions (~6,561× acceleration) for 10,000 dimensional QUBO problems, demonstrating 'Quantum Advantage' for large and dense QUBO problems. Pure QA and QA with the QUBO decomposition method still exhibit limitations in solving large problems due to the current QA hardware limitations (e.g., number of qubits and qubit connectivity). However, we anticipate that QA will eventually reach the efficiency of HQA with the ongoing development of the quantum hardware. Thus, we expect QA to demonstrate true 'Quantum Advantage' in the future.

## Methods

### Definition of a QUBO

QA hardware is designed to efficiently solve combinatorial optimization problems that are formulated with a QUBO matrix, which can be given by[28,29]:

$$y = \sum_{i=1}^{n} \sum_{j=i}^{n} Q_{i,j} x_i x_j \qquad (1)$$

where $Q_{i,j}$ is the $i$-th row and $j$-th column real-number element of the QUBO matrix (**Q**), which is an $n \times n$ Hermitian, i.e., $\mathbf{Q} \in \mathbb{R}^{n \times n}$, and $x_i$ is the $i$-th element of a binary vector $\boldsymbol{x}$ with a length of $n$, i.e., $\boldsymbol{x} \in {0, 1}^n$. $Q_{i,j}$ is often referred to as a linear coefficient for $i = j$ and a quadratic interaction coefficient for $i \neq j$. The objective of QA is to identify the optimal binary vector of a given QUBO, which minimizes the scalar output $y$ as[29]:

$$\boldsymbol{x}^* = \underset{x}{\mathrm{argmin}}\, y \qquad (2)$$

In optimization problems, the linear coefficients correspond to cost or benefit terms associated with individual variables, while the quadratic coefficients represent interaction terms or dependencies between pairs of variables. These coefficients can be learned using machine learning models, such as the factorization machine (FM), trained on datasets containing input structures and their corresponding performance metrics. By mapping these learned coefficients into a QUBO formulation, we effectively represent an energy function of a material system or other real-world optimization problem. This QUBO then describes the optimization space, enabling the identification of the optimal state with the best performance[30,31].

**Methods to Solve a QUBO**

Various methods have been proposed to solve QUBO problems. For our benchmarking study, we consider seven representative methods: QA, hybrid QA (HQA), integer programming (IP), simulated annealing (SA), steepest descent (SD), tabu search (TS), parallel tempering with isoenergetic cluster moves (PT-ICM). Below, we provide a brief introduction to each of the solvers used in solving combinatorial optimization problems:

*Quantum Annealing and Hybrid Quantum Annealing*

QA starts with a superposition state for all qubits, which has the lowest energy state of the initial Hamiltonian ($H_0$). In the annealing process, the system evolves toward the lowest energy state of the final Hamiltonian (also called a problem Hamiltonian, $H_p$) by minimizing the influence of the initial Hamiltonian. The measured state at the end of the annealing is supposed to be the ground state of $H_p$, which can be expressed as the following equation[32,33]:

$$H(t/t_a) = A(t/t_a)H_0 + B(t/t_a)H_p \qquad (3)$$

Here, $t$ is the elapsed annealing time, and $t_a$ is the total annealing time. Equation (3) evolves from $A(t/t_a) = 1$, $B(t/t_a) \approx 0$ at the beginning of the annealing ($t/t_a = 0$) to $A(t/t_a) \approx 0$, $B(t/t_a) = 1$ at the end of the annealing ($t/t_a = 1$). Sufficiently slow evolution from $H_0$ to $H_p$ enables the quantum system to stay at the ground state, which leads to the identification of the optimal solution of a given combinatorial optimization problem[3,34]. We use D-Wave Systems' quantum annealer (Advantage 4.1) to solve the problems using QA, and we set the number of reads for QA to 1,000 with a total annealing time of 20 μs. We select the best solution corresponding to the lowest energy state found among 1,000 reads.

The D-Wave Ocean software development kit (SDK, ver. 3.3.0) provides many useful libraries, which include quantum or classical samplers such as the QA, HQA, SA, SD, and TS. They allow us to solve QUBO problems[22,35,36]. We employ these samplers, which are implemented in the D-wave Ocean SDK, for the benchmarking study. Classical or QA solvers often benefit from decomposition algorithms to identify a high-quality solution (i.e., an optimal solution or a good solution close to the global optimum) for large QUBO problems. Hence, the decomposition of a QUBO matrix into sub-QUBOs is very useful when the size of QUBO matrix is larger than the physical volume of a sampler (i.e., QUBO size > physical number of qubits in QA or memory capacity of a classical computer). We employ the QBSolv package implemented in D-wave Ocean SDK for QUBO decomposition. The QBSolv splits a QUBO matrix into smaller QUBO matrices, and each of them is sequentially solved by classical or QA solvers. This algorithm enables us to handle a wide range of complex real-world problems[21,22,37]. The size of the decomposed QUBOs is set to 30 unless otherwise specified. HQA (Leap Hybrid solver), developed by D-Wave systems, also decomposes large QUBO into smaller subproblems well-suited for QA's QPU, and then aggregates the results[27,38]. The detailed algorithm of HQA, however, is not publicly released. We utilize a D-Wave sampler (dwave-system 1.4.0) for SA, SD, and TS with a specified number of reads (1,000) and default settings for other parameters. Furthermore, we employ D-Wave hybrid framework for PT-ICM.

*Integer Programming*
IP uses branch-and-bound, cutting planes, and other methods to search the solution space for optimal integer decisions and prove global optimality within a tolerance (gap). We use Gurobi (version 10.0.2) [39] for benchmarking with the default settings (0.1% global optimality gap) plus a two-hour time limit and 240 GB software memory limit per optimization problem. The benchmark QUBO problem is implemented in the Pyomo modeling environment (version 6.6.2) [40]. We also experimented with a large gap and observed the first identified integer solution often had a poor objective function value. These results are not further reported for brevity.

*Simulated Annealing*
SA, which is inspired by the annealing process in metallurgy, is a probabilistic optimization algorithm designed to approximate a global optimum of a given objective function. It is considered a metaheuristic method, which can be applied to a wide range of optimization problems [41,42]. In SA, temperature and cooling schedule are major factors that determine how extensively the algorithm explores the solution space [43]. This algorithm often identifies near-optimal solutions but cannot guarantee that local or global optimality conditions are satisfied. For SA, the hyperparameters are configured as follows: 1,000 reads, 1,000 sweeps, a 'random' initial state generation, and a 'geometric' temperature schedule.

*Steepest Descent*
SD operates by employing variable flips to reduce the energy of a given QUBO through local minimization computations rather than relying on a calculated gradient in a traditional gradient descent algorithm [44]. This algorithm is computationally inexpensive and beneficial for local refinement; thus, it can be used to search for local optima. In our benchmarking study, SD utilizes hyperparameters set to 1,000 reads and a 'random' strategy for initial state generation.

*Tabu Search*
TS is designed to solve combinatorial and discrete optimization problems by using memory to guide the search for better solutions, as introduced by Glover [45]. This algorithm can escape already visited local minima by remembering those points (called 'Tabu List' to keep track of moves during the search), aiming to identify high-quality solutions in a large solution space. This algorithm works well for combinatorial optimization problems with small search spaces. However, it can be hard to evaluate neighboring solutions and to maintain and update the Tabu List with increasing problem sizes. The hyperparameter settings for TS are as follows: 1,000 reads, a timeout of 100 ms, and 'random' initial state generation.

*Parallel Tempering with Isoenergetic Cluster Moves (PT-ICM)*
PT-ICM is an advanced Monte Carlo method designed to navigate optimization space, such as QUBO problems [46-48]. PT operates by maintaining multiple replicas of the system at different temperatures and allowing exchanges between replicas based on a Metropolis criterion. This approach helps lower-temperature replicas escape local minima with the aid of higher-temperature replicas. ICM identifies clusters of variables that can flip without changing the system's energy [46]. In this study, the hyperparameters for PT-ICM are set as follows: the number of sweeps is 1,000, the number of replicas is 10, and the number of iterations is 10.

**Benchmarking Problems**

*Real-world problems*

Material optimization is selected to represent real-world problems, with the design of planar multilayers (PMLs) optical film as a testbed for benchmarking. PMLs can be seen in many applications. For example, they have been explored for transparent radiative cooling windows to address global warming by emitting thermal radiation through the atmospheric window (8 μm < λ < 13 μm) [4], while transmitting visible photons. PMLs consist of layers with one of four dielectric materials: silicon dioxide, silicon nitride, aluminum oxide, and titanium dioxide. The configuration of these layers can be expressed as a binary vector, where each layer is assigned a two-digit binary label. Optical characteristics and corresponding figure-of-merit (FOM) of the PML can be calculated by solving Maxwell's equations using the transfer matrix method (TMM). To formulate QUBOs, layer configurations (input binary vectors) and their FOMs (outputs) are used to train the FM model. FM learns the linear and quadratic coefficients, effectively modeling the optimization landscape of the material system. QUBO matrices are then generated using these coefficients[30,31]. PML configurations are randomly generated for training datasets, and their FOMs are calculated using TMM. The resulting QUBO matrices represent real-world materials optimization problems, characterized by highly dense (fully connected) configurations (Fig. S1), which are used for the benchmarking study in Fig. 1.

*Benchmarking problems*

We formulate QUBO matrices with random elements to further systematically explore scalability (Fig. 2 and Fig. 3), following the characteristics of QUBOs from real-world problems, for the benchmarking study as the following:

- Problem size**:** The problem size, corresponding to the length of a binary vector ($n$), varies from 120 to 10,000 (120, 200, 500, 1,000, 1,500, 2,000, 2,500, 3,000, 4,000, 5,000, 6,000, 7,000, 8,000, 9,000 and 10,000).

- Distribution of elements: For each problem size, four QUBO matrices with different distributions of elements are studied. These elements are random numbers with a mean value of 0 and standard deviations of 0.001, 0.01, 0.1, or 1. These distributions reflect the variability observed in QUBO coefficients derived from real-world problems (Table S1). A QUBO configured with elements having a large deviation yields a significant variation in the energy landscape, potentially resulting in high energy barriers that must be overcome to find the ground state.

- Density of matrices: The density of QUBO matrices reflects the proportion of pairwise interactions among variables relative to the maximum possible interactions. Fully connected QUBOs, such as those derived from real-world problems, represent cases where all variables interact with each other. For example, in layered photonic structures, each layer interacts with every other layer, influencing optical responses, which leads to a fully connected QUBO. In contrast, Max-Cut problems typically result in sparse QUBOs, where only a subset of variables (nodes) interact through edges. The maximum number of interaction coefficients (i.e., the number of edges in Max-Cut problems) is $_nC_2$, where $n$ denotes the problem size. The density of a QUBO can be calculated as:

$$\text{density} = \frac{\text{number of interaction coefficients}}{\text{maximum number of interaction coefficients}} \quad (4)$$

For example, a benchmark problem instance (G10) with 800 nodes and 19,176 edges has a density of 6%, calculated as: density = 19,176/319,600 = 0.06. The density of Max-Cut problems can be adjusted by changing the number of edges, with typical instances having densities ranging from 0.02% to 6% (Fig. S1, Table S2). In contrast, real-world problems feature fully connected configurations, corresponding to a density of 100%. QUBOs for this benchmarking study have dense matrices fully filled with real-number elements in the upper triangular part (i.e., fully connected graph nodes, Fig. S2). This configuration aims to approximate real-world optimization problems, which usually requires a dense QUBO matrix[4,28].

**Performance Metrics: Relative Accuracy and Computational Time**

*Relative Accuracy*

For small-scale problems, brute-force search guarantees the identification of the global optimum by evaluating all possible solutions. However, this approach becomes infeasible for large-scale problems due to the exponential growth of the search space. The IP solver, such as Gurobi, utilizes the branch-and-bound method to efficiently explore the solution space and prove global optimality within an optimality gap. However, due to computational limitations or time constraints, IP may struggle to find the global optimum for large-scale problems. To address this challenge in our benchmarking study, we employ a 'Relative Accuracy' metric to compare the relative performance of different solvers. Relative accuracy is defined as the ratio of a solver's objective value to the best objective found across all solvers:

$$\text{Relative Accuracy} = \text{Solution}_{\text{solver}} / \text{Solution}_{\text{best}} \quad (5)$$

This metric provides a way to evaluate the solution quality when the global optimum cannot be definitively found or proven for large-scale problem instances. Note that the best solution is the lowest value among the solutions obtained from all solvers since the solvers are designed to find the lowest energy state (generally negative values for the QUBOs used in this study). The relative accuracies of the solvers are plotted as a function of problem sizes. In Fig. 1, the relative accuracy represents the average value calculated from three different QUBOs that represent material optimization, and in Fig. 2, it represents the average from four different QUBOs with varying standard deviations for each problem size (ranging from 120 to 10,000). Error bars on the plot represent the standard deviation of accuracies calculated from the four different QUBOs for each problem size, relative to the average values. By definition, the relative accuracy is 1.0 when the solver finds a solution with the best-known objective function value (equation 5).

*Computational Time*

Computational time is another important factor in determining the solvers' performance. Combinatorial optimization problems are considered NP-hard, so increasing problem sizes can lead to an explosion of search space, posing challenges in optimization processes. We measure the computational time dedicated solely to solving given problems, excluding problem reading time, queue time, or communication time between the local computer and quantum annealer. This is consistent with other benchmarking studies[17,18]. For problems solved on D-Wave systems' QPU for QA, the execution time includes programming and sampling times (anneal, readout, and delay time). QPU access time is calculated for all of them after programmed anneal-read cycles, corresponding to the time charged to users in their allocations, which is used as the computational time for QA and HQA. Classical solvers (SA, SD, TS, and PT-ICM) run on a workstation (AMD

Ryzen Threadripper PRO 3975WX @ 3.5 GHz processor with 32 cores and 32GB of RAM), and IP (Gurobi) run on a cluster node (an Intel(R) Xeon(R) CPU E5-2680 v3 @ 2.50GHz processor with 24 cores and 256 GB of RAM). Problem reading time can be significant when the problem size is large, but it is excluded from the computational time consideration. We measure the time solely taken to solve given problems with classical solvers. In Fig. 1b and Fig. 3, the solution time for classical and quantum solvers is presented as a function of problem sizes. Note that a QUBO problem is NP-hard[49]. Evaluating the energy of a given solution has a computational cost of $O(n^2)$, where $n$ (= problem size) is the number of variables. The number of reads or sweeps does not scale with $n$, but the cost for each sweep scales as $O(n)$ for SA. Consequently, the theoretical time complexities of the classical solvers are known as $O(n^3)$ for SA[50], $O(n^2)$ for SD[51], and $O(n^2)$ for TS[52]. On the other hand, the theoretical time complexity of the quantum solvers can be considered constant.

# Data availability
All data generated and analyzed during the study are available from the corresponding author upon reasonable request.

# Code availability
The codes used for generating and analyzing data are available from the corresponding author upon reasonable request.

# Acknowledgements
This research used resources of the Oak Ridge Leadership Computing Facility at the Oak Ridge National Laboratory, which is supported by the Office of Science of the U.S. Department of Energy under Contract No. DE-AC05-00OR22725. This research was supported by the Quantum Computing Based on Quantum Advantage Challenge Research (RS-2023-00255442) through the National Research Foundation of Korea (NRF) funded by the Korean Government (Ministry of Science and ICT(MSIT)).

# Author information
Authors and Affiliations
**Department of Aerospace and Mechanical Engineering, University of Notre Dame; Notre Dame, Indiana 46556, United States.**
Seongmin Kim & Tengfei Luo

**Department of Electronic Engineering, Kyung Hee University; Yongin-Si, Gyeonggi-do 17104, Republic of Korea.**
Sangwoo Ahn & Eungkyu Lee

**Department of Chemical and Biomolecular Engineering, University of Notre Dame; Notre Dame, Indiana 46556, United States.**

Alexander Dowling

**National Center for Computational Sciences, Oak Ridge National Laboratory, Oak Ridge, Tennessee 37830, United States.**
Seongmin Kim & In-Saeng Suh

**Contributions**
S.K., A.D., E.L., and T.L. conceived the idea. S.K. and S.A. performed benchmarking studies to generate data. A.D. and S.K. implemented the IP benchmark. S.K. analyzed the data with advice from I.S., A.D., E.L., and T.L. All authors discussed the results and contributed to the writing of the manuscript.

**Corresponding authors**
Correspondence to Alexander W. Dowling, Eungkyu Lee, or Tengfei Luo.

# Ethics declarations
**Competing Interests**
The authors declare no competing interests.

# Figures

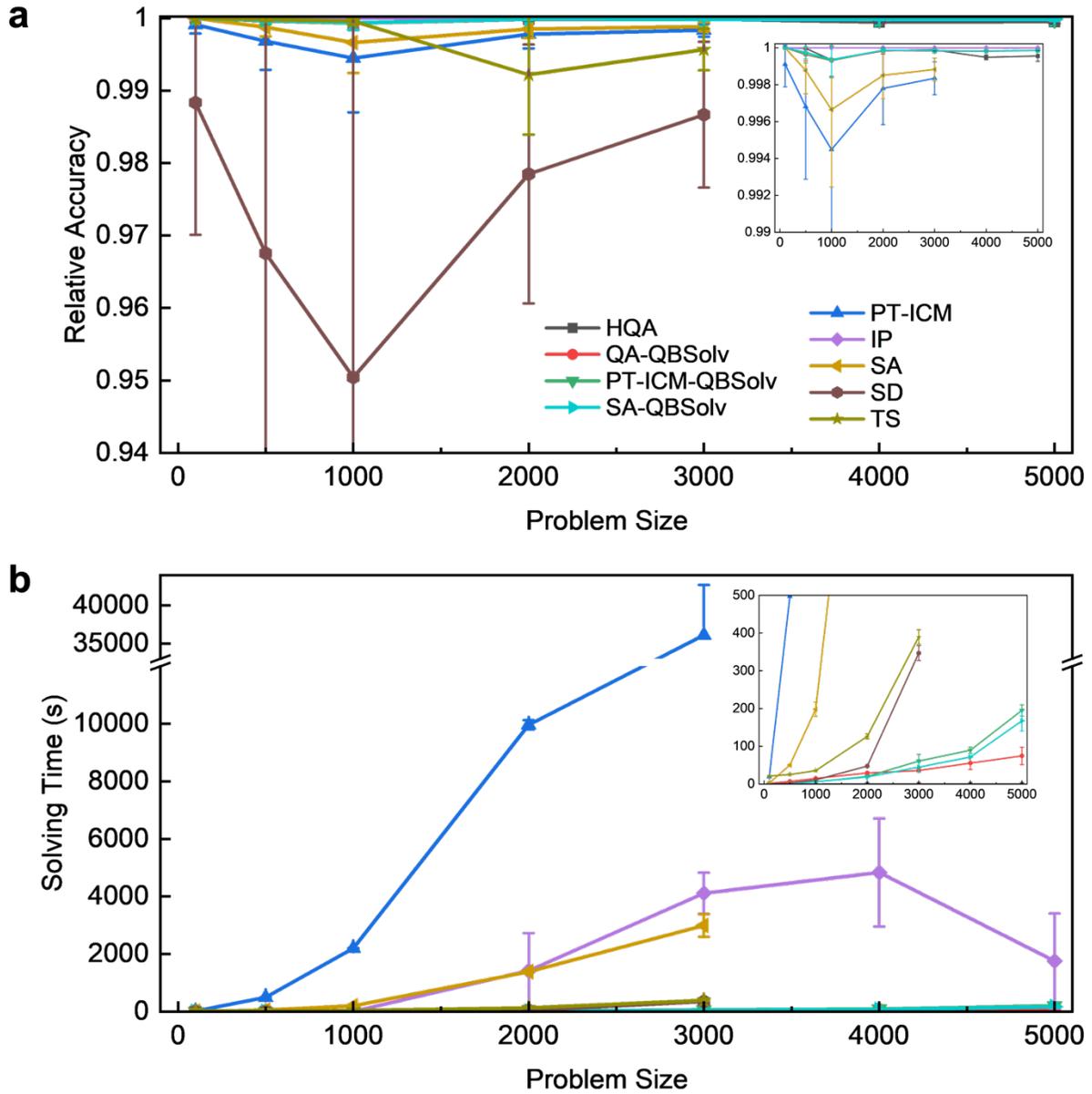

**Fig. 1. Performance analysis of classical (IP, SA, SD, TS, PT-ICM, SA–QBSolv, and PT-ICM–QBSolv) and quantum (QA–QBSolv, and HQA) solvers on QUBO problems representing real-world optimization tasks in material science.** (**a**) Relative accuracy and (**b**) solving time of the solvers.

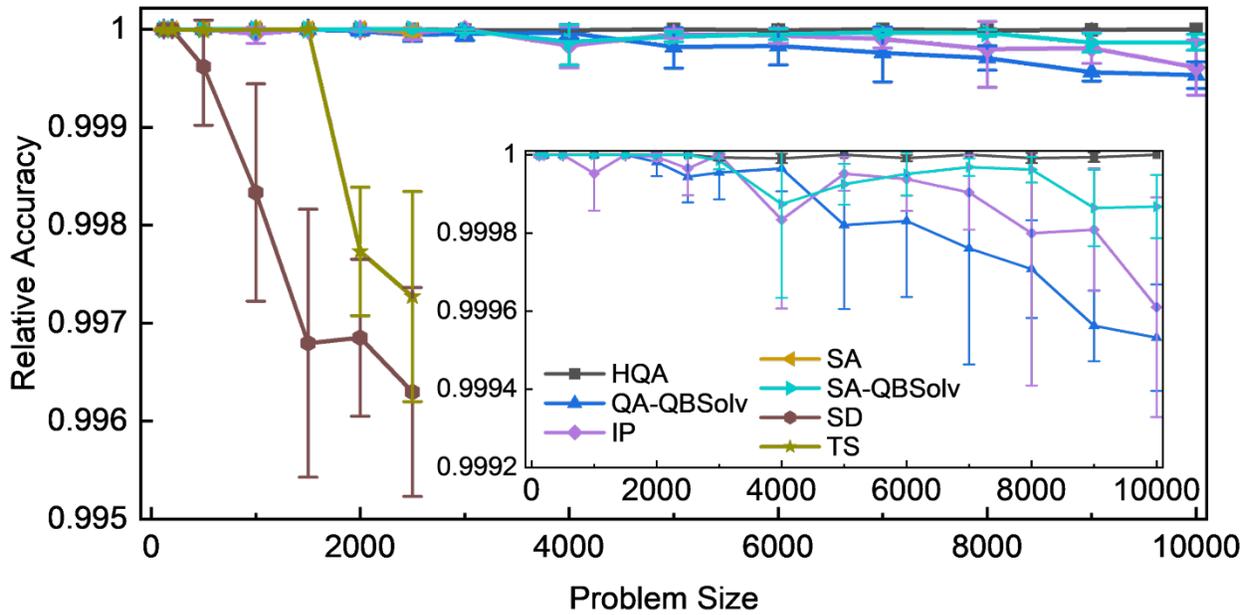

**Fig. 2. The relative accuracy of the classical (IP, SA, SD, TS, and SA–QBSolv) and quantum (QA–QBSolv, and HQA) solvers for given QUBO problems.** HQA is the best solver for finding the highest-quality solution for all problem sizes.

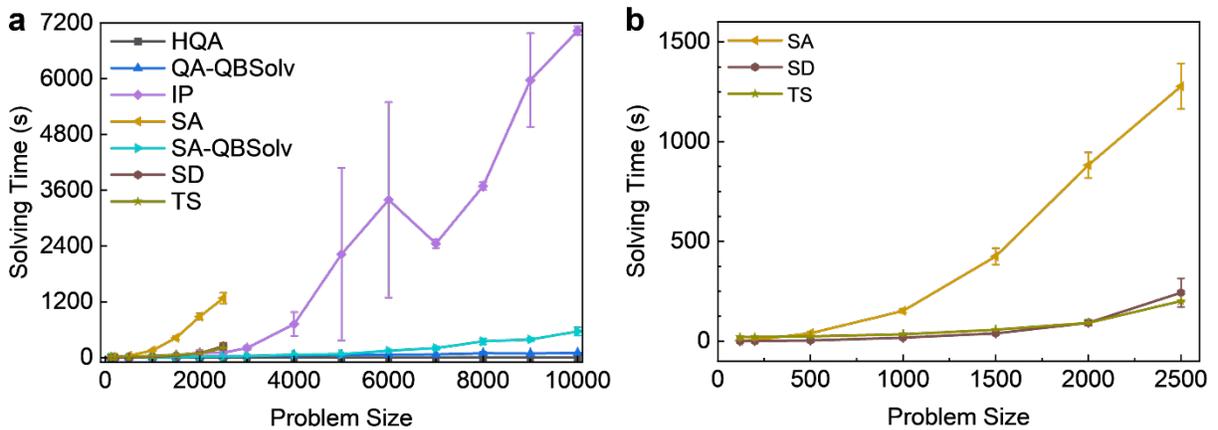

**Fig. 3. Solving time of the solvers for given QUBO problems.** The solving time of (**a**) the classical and quantum solvers and (**b**) the classical solvers (SA, SD, and TS) for small QUBO problems. Quantum solvers do not scale in solving time as the problem size increases, which is a great advantage over classical counterparts.

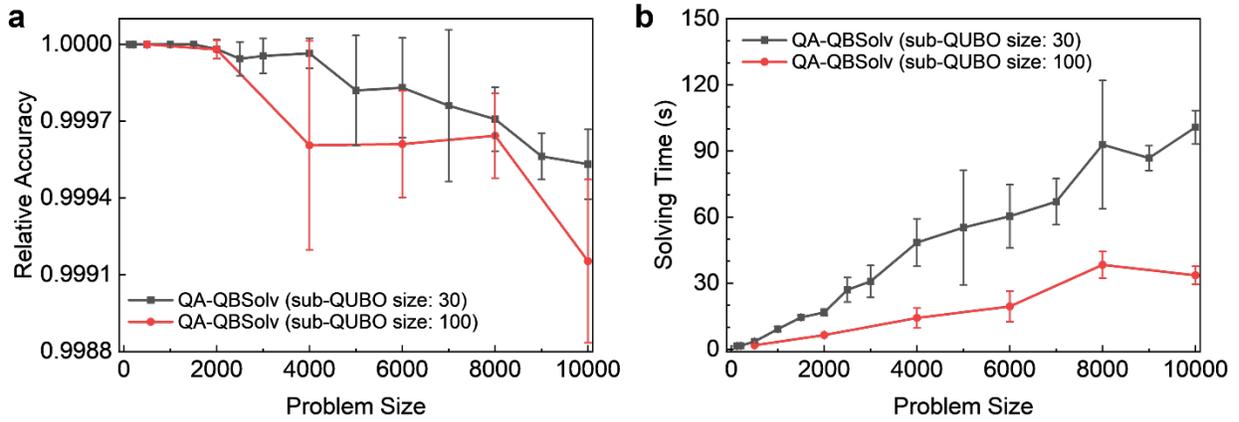

**Fig. 4. Performance of the QA–QBSolv solver with different decomposition sizes.** (**a**) Relative accuracy and (**b**) Solving time of the QA–QBSolv solver for given QUBO problems with different sub-QUBO sizes.

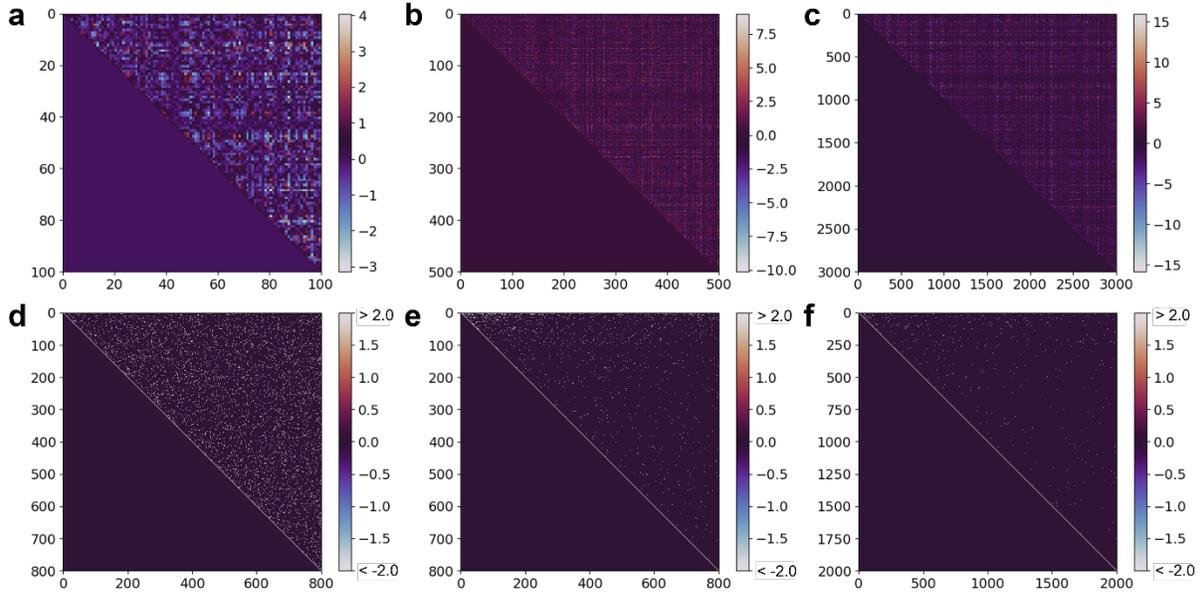

**Fig. S1. Comparison of QUBO matrices for real-world optimization and Max-Cut problems.** (**a–c**) QUBO matrices representing the optimization of planar multilayered structures (PMLs) with problem sizes of (**a**) 100, (**b**) 500, and (**c**) 3,000. The dense configurations of these matrices reflect the fully connected nature of interactions in material optimization problems. (**d–f**) QUBO matrices derived from Max-Cut problem instances in the G-set[S1]: (**d**) G5, (**e**) G15, and (**f**) G40. These matrices exhibit sparse configurations, with relatively few pairwise interactions compared to their maximum possible connections.

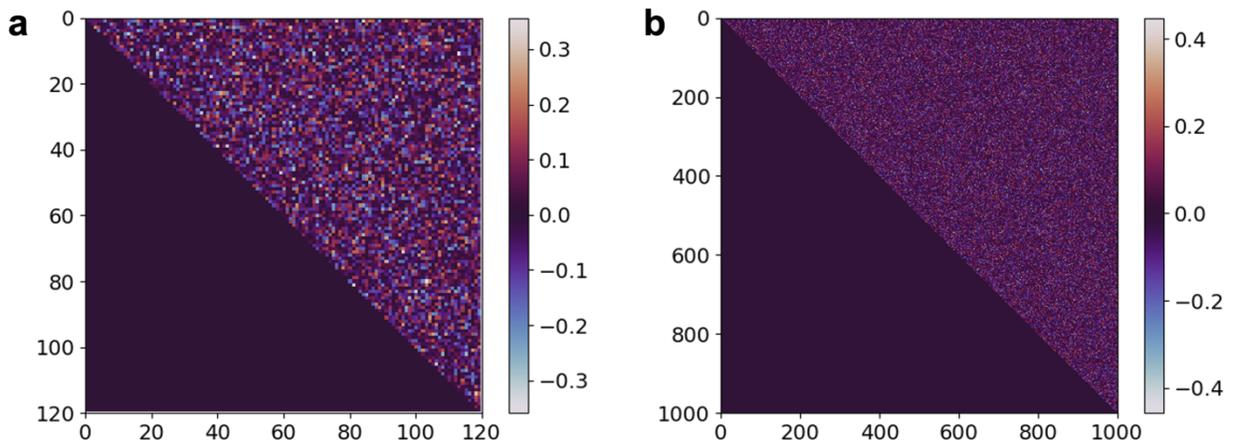

**Fig. S2. Example QUBO matrices.** The size of the given QUBO problems is (**a**) 120 and (**b**) 1,000 with a standard deviation of 0.1.

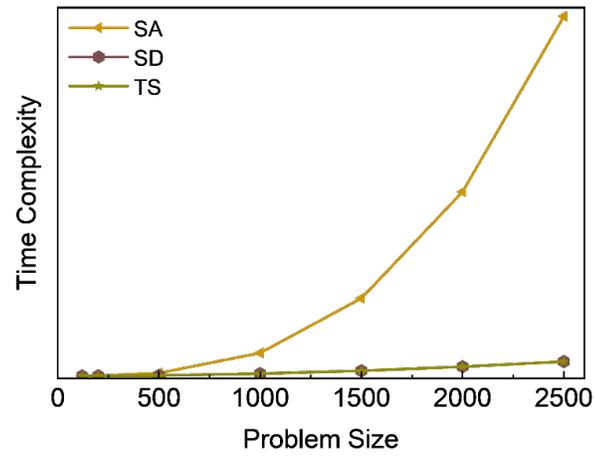

**Fig. S3. Time complexity of simulated annealing (SA), steepest descent (SD), and tabu search (TS).** This plot is from calculation results based on the theoretical time complexity (see 2-4-2. Computational Time in the main text), so it does not have metrics. The plot agrees well with the solving time plot depicted in Fig. 2b.

**Table S1. Statistical properties of QUBO coefficients for real-world optimization problems.** The table summarizes the average (avg) and standard deviation (std) of QUBO coefficients across different problem sizes ($n$). The average values of the coefficients are close to zero, and the standard deviation ranges from 0.2 to 2.

| $n$ | 50 | 100 | 200 | 500 | 1000 | 3000 | 5000 | 10000 |
|---|---|---|---|---|---|---|---|---|
| **avg** | 0.0025 | -0.0014 | 0.0003 | -0.0004 | 0.0001 | 0.0016 | 0.0012 | 0.0008 |
| **std** | 0.2491 | 0.7440 | 0.8083 | 1.3319 | 1.5090 | 1.9519 | 2.0372 | 2.0706 |

**Table S2. Density of Max-Cut problem instances.** These instances feature sparse QUBO matrices with a density lower than 6%.

| Instances | # Nodes | # Edges | # Maximum Edges | Density (%) |
|---|---|---|---|---|
| G5 | 800 | 19,176 | 319,600 | 6.0000 |
| G10 | 800 | 19,176 | 319,600 | 6.0000 |
| G15 | 800 | 4,661 | 319,600 | 1.4583 |
| G20 | 800 | 4,672 | 319,600 | 1.4618 |
| G30 | 2,000 | 19,900 | 1,999,000 | 0.9954 |
| G40 | 2,000 | 11,766 | 1,999,000 | 0.5885 |
| G50 | 3,000 | 6,000 | 4,498,500 | 0.1333 |
| G55 | 5,000 | 12,498 | 12,497,500 | 0.1000 |
| G60 | 7,000 | 17,148 | 24,496,500 | 0.0700 |
| G70 | 10,000 | 9,999 | 49,995,000 | 0.0200 |